\documentstyle[12pt,psfig,epsfig,amsmath,amssymb,bbold]{article}

\def\slash#1{\setbox0=\hbox{$#1$}               
   \dimen0=\wd0                                 
   \setbox1=\hbox{/} \dimen1=\wd1               
   \ifdim\dimen0>\dimen1                        
      \rlap{\hbox to \dimen0{\hfil/\hfil}}      
      #1                                        
   \else                                        
      \rlap{\hbox to \dimen1{\hfil$#1$\hfil}}   
      /                                         
   \fi}                                         %

\def\be{\begin{eqnarray}}
\def\ee{\end{eqnarray}}

\def\BA{\begin{eqnarray}}

\def\BE{\begin{equation}}

\def\EA{\end{eqnarray}}

\def\EE{\end{equation}}

\def\gtsim{\lower-0.45ex\hbox{$>$}\kern-0.77em\lower0.55ex\hbox{$\sim$}}

\def\ltsim{\lower-0.45ex\hbox{$<$}\kern-0.77em\lower0.55ex\hbox{$\sim$}}

\newcommand{\bea}{\begin{eqnarray}}
\newcommand{\eea}{\end{eqnarray}}       

\begin{document}


\title{\bf The Spectrum of the Dirac Operator in the Linear
  $\sigma$--Model with Quarks}

\author{Thomas Spitzenberg$^{a}$,
\\[4mm]
\it
    Institut f\"ur Kernphysik, Universit\"at Mainz,
\\
\it
    Johann-Joachim-Becher-Weg 45, 55099 Mainz, Germany
\\[4mm]
Kai Schwenzer$^{b}$, Hans--J\"urgen Pirner$^{c}$
\\[4mm]
\it
    Institut f\"ur Theoretische Physik, Universit\"at Heidelberg,
\\
\it
    Philosophenweg 19, 69120 Heidelberg, Germany
\\[4mm]
\small{(a) spitzenb@kph.uni-mainz.de, (b)
  kai@tphys.uni-heidelberg.de,} \\ 
\small{(c) pir@tphys.uni-heidelberg.de}}

\maketitle

\begin{abstract}
  We derive the spectrum of the Dirac operator for the linear
  $\sigma$--model with quarks in the large
  $N_c$ approximation using renormalization group flow equations. For
  small eigenvalues, the Banks--Casher relation and the vanishing
  linear term are recovered. We calculate the coefficient of the next
  to leading term and investigate the spectrum beyond the low energy regime.
\end{abstract}


\newpage
\section{Introduction}
The QCD Dirac operator contains all information about the quark
dynamics in QCD. Therefore its spectrum is of high interest. In the
literature one considers the Euclidean Dirac operator which reads
$D=\slash\partial+i\,g\slash A$. We will use Euclidean space time
throughout this work. Since the Dirac operator is antihermitean, its
eigenvalues are purely imaginary. With the eigenvalue equation
\begin{equation}
D\left| \psi_k \rangle =i\lambda_k\right|\psi_k\rangle 
\end{equation}
one defines the spectral density
\begin{equation}
\label{eq:QCDspectrum}
\rho(\lambda)=\Bigl\langle\sum_k\delta\left(
  \lambda-\lambda_k\right)\Bigr\rangle \; ,
\end{equation}
where the averaging is over the gluon background. The first discovery
about the spectral density has been made by Banks and Casher
\cite{Banks:1980yr} who connect the order
parameter of spontaneous chiral symmetry breaking, i.e. the chiral
condensate, to the QCD spectrum at zero eigenvalues. Since then,
different approaches have led to a steady progress in this field. As
the dynamics at low energies is mainly driven by chiral symmetry the
behaviour of the spectral density in the vicinity of $\lambda=0$ can
be analysed using chiral perturbation theory. The first correction to
the Banks--Casher relation was found by Smilga and Stern
\cite{Smilga:1993in}
\begin{equation}
\label{eq:SmilgaSternresult}
\rho(\lambda)=\frac{V_4\langle \bar q q
  \rangle}{\pi}+V_4\frac{\langle\bar q q \rangle^2\left(
    N_f^2-4\right)}{32\pi^2N_f f_\pi^4}\left| \lambda
\right|+O(\lambda^2) \; ,
\end{equation}
where $V_4$ denotes the Euclidean 4--volume. The constant term is the
Banks--Casher result, whereas the linear term vanishes in the case $N_f=2$. \\ 
Chiral random matrix theory considers the so called microscopic
spectral density, in which all eigenvalues are rescaled according to
the average spacing of small eigenvalues. It successfully describes the
Dirac spectrum in the extreme infrared regime, where the spectral
density is completely determined by the global symmetries of the Dirac
operator (see e.g. \cite{Verbaarschot:2000dy}). Universal properties
of the microscopic spectral density have been derived and identified
in lattice calculations \cite{Berbenni-Bitsch:1998tx,Damgaard:2001ep}. Furthermore
partially quenched chiral perturbation theory combines the results of
chiral perturbation theory and chiral random matrix theory
\cite{Osborn:1999qb}.\\
The Dirac spectrum for small eigenvalues is driven by chiral symmetry
which determines hadron dynamics for small momenta. Thus chiral
perturbation theory and chiral random matrix theory can be successful
in describing the Dirac spectrum.\\
The same holds for the Nambu--Jona-Lasinio (NJL) model which serves as an
effective model for chiral symmetry. In this work we consider it in its
bosonized form given by the linear $\sigma$-model. The QCD
Dirac spectrum and the spectrum for the linear $\sigma$--model which we
analyze in this paper should coincide in the regime, where the linear
$\sigma$--model is a good effective description for QCD, that is for 
\begin{equation}
\lambda < \Lambda\;.
\end{equation}
Here $\Lambda \approx 1$ GeV defines the momentum scale below which a
hadronic description of QCD may be useful.

\section{RG flow in the linear $\sigma$--model}

The Lagrangian of the linear $\sigma$--model in Euclidean spacetime is
\begin{equation}
\label{eq:lagrangian}
{\cal L}=\bar q \left[ Z_q\slash\partial+g \left(
    \sigma+i\vec\tau\vec\pi\gamma_5 \right) \right] q +
\frac{1}{2}Z_\Phi \left( (\partial_\mu \sigma)^2+(\partial_\mu
  \vec\pi)^2 \right)+V(\sigma,\vec\pi,m_q) \; .
\end{equation}
The fermion fields describe constituent quarks which appear in two
flavours and three colours. The bosonic potential 
\begin{equation}
  V(\sigma,\vec\pi,m_q)=U(\Phi^2)-c \, \sigma
\end{equation}
contains a chiral symmetric part $U(\Phi^2)$ depending on the
$O(4)$-symmetric representation $\Phi\!=\!(\sigma,\vec\pi)$ and a
linear symmetry breaking term $c \, \sigma$. For later purposes we
explicitly stress the dependence of the meson potential
$V(\sigma,\vec\pi,m_q)$ on the current quark mass $m_q$. At the
ultraviolet scale, bosonization of the NJL model fixes the form of the
meson potential $V_\Lambda(\sigma,\vec\pi,m_q)$ to be
\begin{equation}
\label{eq:HiggsPotential}
V_\Lambda(\sigma,\vec\pi,m_q)=\frac{m_\Lambda^2}{2} \Phi^2-c \, \sigma \; .
\end{equation}
The linear symmetry breaking term $c$ is related to the current quark mass
$m_q$ by
\begin{equation}
\label{eq:mass2linear}
c=\frac{m_\Lambda^2}{g_\Lambda}\,m_q \; .
\end{equation}
We use renormalization group (RG) flow equations for the linear
$\sigma$-model with quarks in the Schwinger proper time
formalism. These flow equations have been discussed in various
papers. For a general review on RG flow equations see \cite{RGreview}
and for the Schwinger proper time approach used here see
\cite{Schaefer:1999em}. The evolution equations provide a
framework to study nonperturbative theories by
integrating out smaller and smaller momentum shells above a limiting
infrared cutoff $k$. This cutoff enters in the form of a smooth
regulator function $f_k(\tau)$ in the Schwinger proper time
representation of the logarithm containing one-loop fluctuations. We
consider the set of regulator functions given by 
\begin{equation}
\label{eq:cutoffs}
  f^{(n)}(\tau k^2)=\sum_{j=0}^n \frac{(\tau k^2)^j}{j!} e^{-\tau k^2}
  \; .
\end{equation}
The flow equations which we present in this paper are obtained with
$n=2$. In order to check the cutoff dependence of our results we use
regulator functions with different n. Generalization of the equations
is straight forward.\\
Note, the parameter $c$ is scale independent, since a linear symmetry
breaking term does not evolve under RG flow
\cite{Zinn-Justin}. Therefore only the chiral symmetric part
$U_k(\Phi^2)$ of the meson potential
$V_k(\sigma,\vec\pi,m_q)=U_k(\Phi^2)-c\sigma$ evolves under flow of
the IR cutoff scale $k$. \\
In this paper we restrict ourselves to the flow
equations in the large $N_c$ approximation, which
have been considered recently \cite{Meyer:2001zp} and whose results have
proven equivalent to the standard selfconsistent NJL large $N_c$
approach \cite{Klevansky:1992qe,Ripka}. They are particularly simple and read
\begin{align}
\label{eq:large-nc-pot}
  k\frac{\partial U_k}{\partial k}
  &= -\frac{N_f N_c}{8\pi^2} \; \frac{k^6}{k^2\!+\!\Phi^2} \; , \\  
\label{eq:large-nc-z}
  k\frac{\partial Z_{\Phi k}}{\partial k} &= -\frac{N_f N_c}{4\pi^2} \,
  \frac{k^6}{(k^2\!+\!\Phi^2)^3} \; .
\end{align}
The couplings $Z_q$ and $g$ in eq. (\ref{eq:lagrangian}) do not evolve in large $N_c$ and have been set to 1
for convenience. \\
An additional flow equation which determines the chiral condensate can
be obtained by differentiation of the partition function $Z$ with respect
to an appropriately introduced source term $\Delta \bar q q$ which
probes the chiral condensate
\begin{equation}
  \langle \bar q q\rangle=\left. \frac{\partial}{\partial \Delta}\log
  Z(\Delta)\right|_{\Delta=0}\;.
\end{equation}
This flow equation is derived analogously to the other flow equations by
introduction of the Schwinger proper time cutoff in $Z$ and reads
\begin{equation}
\label{eq:Kondensat-Flussgleichung}
k \frac{\partial \langle \bar q q \rangle_k}{\partial k}=\frac{N_f N_c}{4 \pi^2} \frac{k^6 \,
  \sigma}{(k^2+\Phi^2)^2} \; . 
\end{equation}
In large $N_c$ all flow equations can be solved analytically. The
initial conditions and solutions are discussed in \cite{Meyer:2001zp}. Denoting $\Phi=(\sigma,\vec\pi)$ one finds for the potential (at $k=0$)
\begin{equation}
\label{eq:potsol}
V(\sigma,\vec\pi,m_q)=V_\Lambda(\sigma,\vec\pi,m_q)+\frac{N_c}{8\pi^2}\left( \Phi^4\log\left(
    \frac{\Lambda^2+\Phi^2}{\Phi^2}\right)-\Phi^2\Lambda^2
\right)\; .
\end{equation}
The minimum of this potential yields the vacuum expectation value $\Phi_0$ of the bosonic
field which is nonvanishing due to the spontaneous and explicit
breaking of chiral symmetry. 
The physical value for the quark condensate is obtained as $\langle \bar q q
\rangle (\Phi_0)$. Its flow is shown in Fig. 1.
 \begin{figure}[h]
\begin{center}
\begin{raggedright}
\epsfig{file=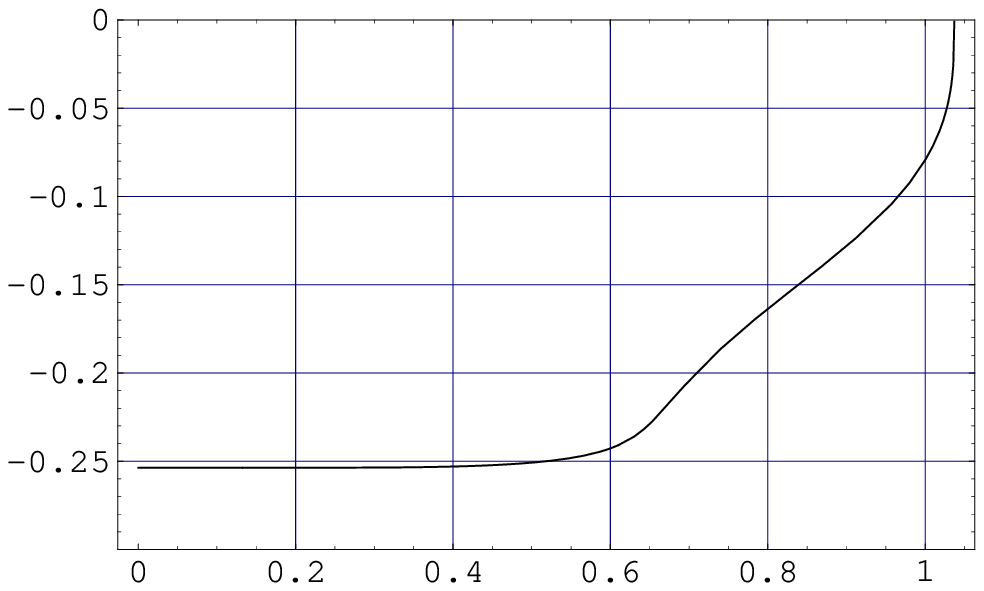}
\end{raggedright} 
\centerline{$k \; [\mbox{GeV}]$}
\flushleft \vspace{-4.0cm} \hspace{-0.2cm} {$\langle \bar q q
  \rangle^{\frac{1}{3}} \; [\mbox{GeV}]$}
\vspace{+3.4cm} 
\caption{\label{fig:condensate}The evolution of the chiral quark
  condensate as a function of the renormalization scale $k$ with an
  explicit symmetry breaking due to a current quark mass of $m_q=7$ MeV.
}
\end{center}
\end{figure} \\
We obtain an infrared value $\langle \bar q q \rangle^\frac{1}{3}_{RG}=253.6$
MeV, which compares well to the value obtained from the
Gell-Mann--Oakes--Renner relation $\langle \bar q q
\rangle^\frac{1}{3}_{GOR}=255.2$ MeV. This result is not surprising (cf. \cite{Ripka}) since the initial conditions of the flow equations have been adjusted to reproduce $f_\pi = 93$ MeV and $M_q = 320$ MeV at $k=0$ using a
current quark mass $m_q = 7$ MeV at the ultraviolet scale.

\section{Extracting the Dirac operator spectrum from the flow equations}
 
\begin{figure}[h]
\label{fig:epsilon}
\begin{center}
\begin{raggedright}
\hspace{1cm}
\epsfig{file=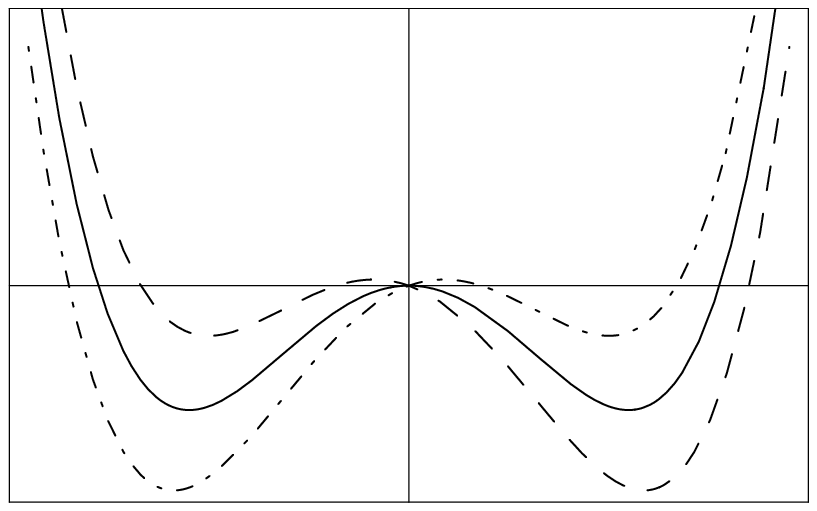}
\hspace{-1cm}
\end{raggedright} 
\centerline{\hspace*{1.1cm}$\sigma$}
\flushleft \vspace{-3.6cm} \hspace{.2cm}
$\text{Re}\left(V(\sigma,0,\epsilon)\right)$
\vspace{+3.0cm} 
\hspace{1.5 cm}
\caption{Projection of the real part of the meson potential depending
  on $\sigma$. The sign of the explicit symmetry breaking term induces
  the choice of the global minimum and hence the vacuum expectation
  value. The solid curve is the potential without explicit symmetry
  breaking. The dashed and dashed dotted curves result from
  $x=+\epsilon$ and $x=-\epsilon$ respectively.}
\end{center}
\end{figure}

\begin{figure}[h!]
\label{fig:spektraldichtegross}
\begin{center}
\begin{raggedright}
\hspace{1cm}
\epsfig{file=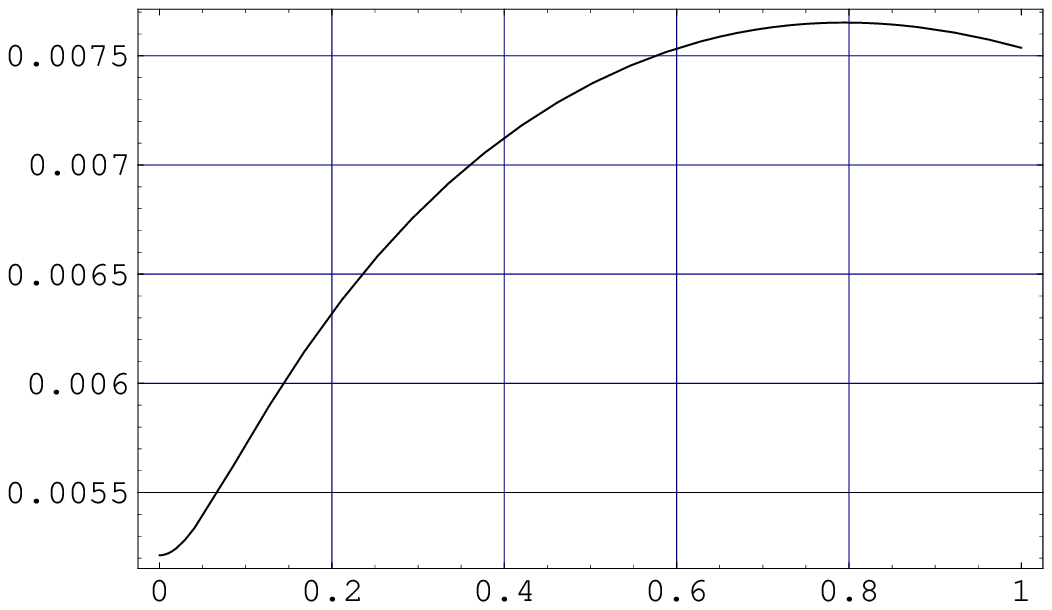}
\hspace{-1cm}
\end{raggedright} 
\centerline{$\lambda \; [\mbox{GeV}]$}
\flushleft \vspace{-4.4cm} \hspace{-0.2cm}{$V_4^{-1}\rho \; [\mbox{GeV}^3]$}
\vspace{+3.8cm} 
\hspace{.6cm}
\caption{\label{fig:lambda-g}The spectral density $\rho(\lambda)$ of the Dirac
  operator $\tilde D$ computed in the linear $\sigma$-model with
  quarks. Note the suppressed zero on the scale for $V_4^{-1} \rho$.
}
\end{center}
\end{figure}

While the QCD--Dirac operator $D$ is antihermitean, the Dirac operator in
the linear $\sigma$--model
\begin{equation}
\label{eq:lsmoperator}
\tilde{D} \!=\! \slash{\partial}+g(\sigma+i \vec\tau \vec\pi \gamma_5)
\end{equation}
can have complex eigenvalues. In the flow
equations we have neglected the imaginary part of the fermion
determinant, which is related to anomalous processes
\cite{Ripka}. In this approximation the eigenvalues of the Dirac operator in the linear $\sigma$-model are
purely imaginary in coincidence with the QCD spectrum. For general complex
eigenvalues $z\!=\!x\!+\!iy$ it is useful to consider the two
dimensional spectral density
\cite{Stephanov:1997cj,efetov}
\begin{equation}
\label{eq:complex-spectral-density}
P(x,y)=\Biggl\langle\,\sum_k\delta(x-x_k) \, \delta(y-y_k) \Biggr
\rangle_\Phi \; ,
\end{equation}
which can be obtained from the complex resolvent \cite{Stephanov:1997cj}
\begin{align}
\label{eq:Resolvent}
G(z) \equiv& \Biggl\langle
Tr\frac{1}{z-\tilde{D}}\Biggr\rangle_\Phi
 = \Biggl\langle Tr \frac{\partial}{\partial_z}\log \left(z-\tilde{D}\right)\Biggr\rangle_\Phi
 =\Biggl\langle
\frac{\partial}{\partial_z}\log\;\text{Det}
\left(z-\tilde{D}\right)\Biggr\rangle_\Phi \nonumber \\
 =& \Biggl\langle\frac{\partial}{\partial_z}\log\int D q \,D \bar q \exp \left[ -\int d^4x \Bigl(
  \bar q \left(z-\tilde{D}\right)q \Bigr) \right]
\Biggr \rangle_\Phi \;.
\end{align}
The complex eigenvalue $z$ is the generator in this resolvent, which
in our low energy effective theory involves an average over the meson
fields denoted by $\langle ...\rangle_\Phi$.
In large $N_c$ approximation there are no meson
fluctuations. Therefore the averaging over the meson fields can be
interchanged with the logarithm in the last line and one finds
\begin{equation}
\label{eq:r2c}
G(z)=-V_4\langle \bar q q\rangle(z)\;.
\end{equation}
The connection between $G(z)$ and $P(x,y)$ can be derived with eq. (\ref{eq:complex-spectral-density})
\begin{align}
\label{eq:G2P}
G(z)&=\Biggl\langle
Tr\frac{1}{z-\tilde{D}}\Biggr\rangle_\Phi \nonumber \\
&=\Biggl\langle\int\;dx^\prime\;dy^\prime\frac{\sum_k \delta(x^\prime-x_k)
  \delta(y^\prime-y_k)}{x-x^\prime+i(y-y^\prime)}\Biggr\rangle_\Phi
\nonumber \\  
&=\int\;
dx^\prime\;dy^\prime\frac{P(x^\prime,y^\prime)}{x-x^\prime+i(y-y^\prime)} \;.
\end{align}
Combining eqs. (\ref{eq:r2c}) and (\ref{eq:G2P}) yields
\begin{equation}
\label{eq:P2qbarq}
G(x,y)=-V_4\langle \bar q q \rangle=\int dx^\prime\int
dy^\prime\frac{P\left(x^\prime,y^\prime\right)}{x-x^\prime+i(y-y^\prime)}\;.
\end{equation}
This relation can be inverted \cite{Stephanov:1997cj} and gives
\begin{equation}
\label{eq:P}
P(x,y)=-\frac{V_4}{\pi}\frac{\partial}{\partial z^\star}\langle\bar q
q \rangle (z)\;, 
\end{equation}
where 
\begin{equation}
\label{eq:def_zstar}
\frac{\partial}{\partial z^\star}=\frac{1}{2}\left(
  \frac{\partial}{\partial x}+i\frac{\partial}{\partial y} \right)\;.
\end{equation}
The complex eigenvalue $z$ enters in the resolvent
eq. (\ref{eq:Resolvent}) just as an explicit quark mass. Therefore no
new flow equations have to be evaluated. After bosonization $z$
appears in a linear symmetry breaking term in the meson potential via
eq. (\ref{eq:mass2linear}) with $m_q=z$.\\ 
As the resulting meson potential $V_k(\sigma,\vec\pi,z)$ becomes
complex, we must
comment on the meaning of $\Phi_{0,k}(z)$. Instead of restricting
ourselves to positive real values $\Phi_{0,k}$, we now have to allow
also complex fields $\Phi_{0,k}(z)$. Instead of finding a real minimum
of the potential $V_k(\sigma,\vec\pi,m_q)$ we have to determine a
complex saddle point, which minimizes the real part of the complex meson
potential $V_k(\sigma,\vec\pi,z)$. \\
From eq. (\ref{eq:P}) it is obvious, that the two dimensional spectral
density is zero if $\langle \bar q q \rangle (z)$ is analytic in
$z$. In presence of the complex quark mass $z$ the
chiral condensate at $k=0$ becomes in the large $N_c$ limit with
eq. (\ref{eq:Kondensat-Flussgleichung})
\begin{equation}
\label{eq:Condensate_Result}
\langle \bar q q \rangle (z)=\frac{N_c}{4\pi^2}\left[
  \frac{-\Phi_0(z)\Lambda^2\left(
  \Lambda^2+2\Phi_0(z)^2\right)}{\Lambda^2+\Phi_0(z)^2}+2\Phi_0(z)^3
  \log \left(1+\frac{\Lambda^2}{\Phi_0(z)^2}\right)\right]\;.
\end{equation}
This expression is analytic if $\Phi_0(z)$ is analytic and
$\text{Re}\left(\Phi_0(z)\right)\neq 0$. As $\Phi_0(z)$ is the bare
pion decay constant in presence of $z$ and therefore minimizes
$\text{Re}\left( V(\sigma,\vec\pi,z)\right)$, the second condition
$\text{Re}\left(\Phi_0(z)\right)\neq 0$ is always true. But the global
minimum of $\text{Re}\left(V(\sigma,\vec\pi,z)\right)$ can switch if $z$ changes
infinitesimally. This is exactly what happens and leads to a
discontinuity in $\Phi_0(z)$. Fig \ref{fig:epsilon} demonstrates
this behaviour.
 \\
For $z=0$ it follows from eq. (\ref{eq:potsol}) in the chiral limit
that with $\Phi_0$ as a global minimum of
$\text{Re}\left(V(\sigma,\vec\pi,z)\right)$ also $-\Phi_0$, $\Phi_0^\star$
and $-\Phi_0^\star$ are global minima of
$\text{Re}\left(V(\sigma,\vec\pi,z)\right)$. Additionally they are saddle points
of $V(\sigma,\vec\pi,z)$. For either $z=x$ or $z=iy$ the global minimum is still twice
degenerate. In case of an infinitesimal $x$ the sign of $x$
chooses the minimum, i.e. $\text{Re}\left(\Phi_0\right)$ changes its
sign under the transformation from $x\!=\!+\epsilon$ to
$x\!=\!-\epsilon$. The same holds for infinitesimal $y$ and for
$\text{Im}\left(\Phi_0\right)$, but
in this case an infinitesimal $y$ only gives rise to an
infinitesimal $\text{Im}\left(\Phi\right)$, consequently no
discontinuity arises in $\Phi(z)$.
We are thus left with the result, that $P(x,y)$ can only be nonzero
for $x=0$. The position of our Dirac spectrum thus coincides with the
QCD spectrum. Using the ansatz $P(x,y)=\delta(x)\rho(y)$ we find from
eqs. (\ref{eq:P2qbarq}), (\ref{eq:P}) and (\ref{eq:def_zstar})
\begin{equation}
\label{eq:spektrum}
\rho(y)=\frac{V_4}{2\pi}\lim_{\epsilon\rightarrow 0} \left[ \langle
  \bar q q\rangle (iy-\epsilon)-\langle \bar q q\rangle (iy+\epsilon)
\right]\;.
\end{equation}
Now let us consider the flow equations for the potential and the chiral
condensate. In the large $N_c$ limit they reveal a simple connection, namely\footnote{To be correct we should have written $\left.\partial_k\langle\bar q q\rangle_k(\sigma(z),\vec\pi)\right|_{\vec\pi=0}=\left.\partial_k\partial_\sigma
U_k(\sigma(z),\vec\pi)\right|_{\vec\pi=0}$. As the expectation value of the pion field vanishes the notation above is unambiguous. }
\begin{equation}
\partial_k\langle\bar q q\rangle_k(\Phi(z))=\partial_k\partial_\Phi
U_k(\Phi(z))\;.
\end{equation}
With the initial conditions $\langle \bar q q \rangle_\Lambda=0$ and
$U_\Lambda=\frac{m_\Lambda^2}{2} \Phi^2$ at $k=\Lambda$ this equation can be
integrated out and yields
\begin{equation}
\label{eq:GORcheck}
\langle\bar q q\rangle\left(\Phi(z)\right)=\partial_\Phi
U\left(\Phi(z)\right)-m_\Lambda^2 \Phi(z) \;.
\end{equation}
Eq. (\ref{eq:spektrum}) gives the spectral density in terms of the
discontinuity of the generalized chiral condensate $\langle \bar q
q\rangle(z)$ across the real axis, which can be calculated from the
saddle point $\Phi_0(z)$ of $V(\Phi,z)$ fulfilling
$\left. \partial_\Phi V(\Phi,z)\right|_{\Phi=\Phi_0}=0$. With
eq. (\ref{eq:GORcheck}) we find in the chiral limit
\begin{align}
\label{eq:rho}
\rho(y)&=\frac{V_4}{2\pi}\lim_{\epsilon\rightarrow 0}m_\Lambda^2\left[ \Phi(i\,y+\epsilon)-\Phi(i\,y-\epsilon) \right] \nonumber \\
&=\frac{m_\Lambda^2V_4}{\pi}\text{Re}\left( \Phi(i y+\epsilon)
\right)\; .
\end{align}
The result is plotted in fig. (\ref{fig:spektraldichtegross}).
\begin{figure}[h]
\begin{center}
\begin{raggedright}
\hspace{1cm}
\epsfig{file=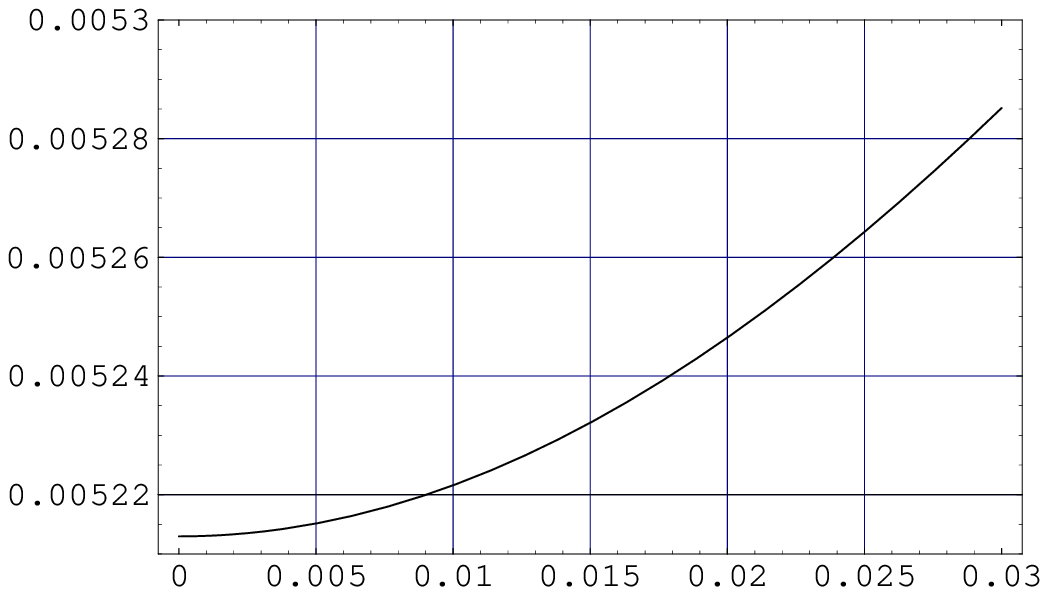}
\hspace{-1cm}
\end{raggedright} 
\centerline{$\lambda \; [\mbox{GeV}]$}
\flushleft \vspace{-4.4cm} \hspace{-0.2cm}{$V_4^{-1}\rho \; [\mbox{GeV}^3]$}
\vspace{+3.8cm} 
\hspace{.6cm}
\caption{\label{fig:lambda-s}Behaviour of the
  spectral density at small eigenvalues, which exhibits a quadratic
  rise due to the absence of a linear contribution in the case $N_f\!=\!2$.
}
\end{center}
\end{figure} \\
Next we derive the behaviour of the spectral density at small
eigenvalues. Therefore we have to evaluate $\Phi_0(iy)$ for small
$y$. The value $\Phi_0(iy):=\Phi_0(0)+\alpha(y)+i\beta(y)$ is
determined by the saddle point equation
\begin{equation}
\left.\partial_\Phi V(\Phi,y)\right|_{\Phi=\Phi_0}=\left.\partial_\Phi
  U(\Phi)\right|_{\Phi=\Phi_0(0)+\alpha(y)+i\beta(y)}-im_\Lambda^2\,y=0\;.
\end{equation}
$\alpha(y)$ and $\beta(y)$ can be expanded in a Taylor series. The
coefficients are obtained from
\begin{equation}
\frac{d^{(n)}}{dy^{(n)}}\left[\left.\partial_\Phi
    U(\Phi)\right|_{\Phi=\Phi(0)+\alpha(y)+i\beta(y)}
    -im_\Lambda^2\,y\right]_{y=0}=0 \quad ,  \quad n=0,1,2 \; .
\end{equation}
We find
\begin{equation}
\Phi_0(y)=\Phi_0(0)+i\frac{m_\Lambda^2}{\left.\partial_\Phi^2
    U(\Phi)\right|_{\Phi=\Phi_0(0)}}y+\frac{1}{2}\frac{m_\Lambda^4
  \left.\partial_\Phi^3 U(\Phi)\right|_{\Phi=\Phi_0(0)}}{\left[
    \left.\partial_\Phi^2 U(\Phi)\right|_{\Phi=\Phi(0)}
  \right]^3}y^2+...
\end{equation}
With eq. (\ref{eq:rho}) the spectral density
\begin{equation}
\label{eq:quadrat}
\rho(y)=\frac{m_\Lambda^2V_4}{\pi}\left[
  \text{Re}\left[\Phi_0(0)\right]+\frac{1}{2}\frac{m_\Lambda^4
    \left.\partial_\Phi^3 U(\Phi)\right|_{\Phi=\Phi_0(0)}}{\left[
      \left.\partial_\Phi^2 U(\Phi)\right|_{\Phi=\Phi_0(0)}
    \right]^3}y^2+O(y^3) \right]
\end{equation}
can be calculated explicitly for small eigenvalues, see
eq. (\ref{eq:potsol}) and (\ref{eq:GORcheck})
\begin{align}
&V_4^{-1}\rho(y)=\frac{\langle \bar q q \rangle}{\pi}+\frac{2\pi^3m_\Lambda^6}{\Phi_0(0)}\\
&\cdot
\frac{12\pi^2m_\Lambda^2\left(\Lambda^2+\Phi_0(0)^2\right)^6
  -N_c\Lambda^6\left(3\Lambda^2-\Phi_0(0)^2\right)\left(\Lambda^2
    +\Phi_0(0)^2\right)^3}{\left[4\pi^2m_\Lambda^2\left(\Lambda^2
      +\Phi_0(0)^2\right)^2 -N_c\Lambda^6\right]^3} \,y^2+O(y^3) \;.\nonumber  
\end{align}
The constant coefficient is the Banks--Casher result
\cite{Banks:1980yr}. Since our computation is done for $N_f=2$ there
is no linear term in agreement with the findings of Smilga and Stern
for $N_f=2$ \cite{Smilga:1993in}. 
Numerically, the quadratic coefficient is 87 MeV
for the choice of parameters given in \cite{Meyer:2001zp}. The
quadratic term in $y$ can be related to the correlation function of
three scalar $\sigma$--quanta. In the numerator of
eq. (\ref{eq:quadrat}) one sees the effective $\sigma^3$ coupling
constant contained in our potential $U(\Phi)$ multiplied by the cube
of the $\sigma$--propagator $\frac{1}{m_\sigma^2}$ at zero
momenta. This result can also be found in the paper of Smilga and Stern \cite{Smilga:1993in}.

\section{The Dirac operator in the strong coupling regime of QCD}

\begin{figure}[h!]
\begin{center}
\begin{raggedright}
\hspace{1cm}
\epsfig{file=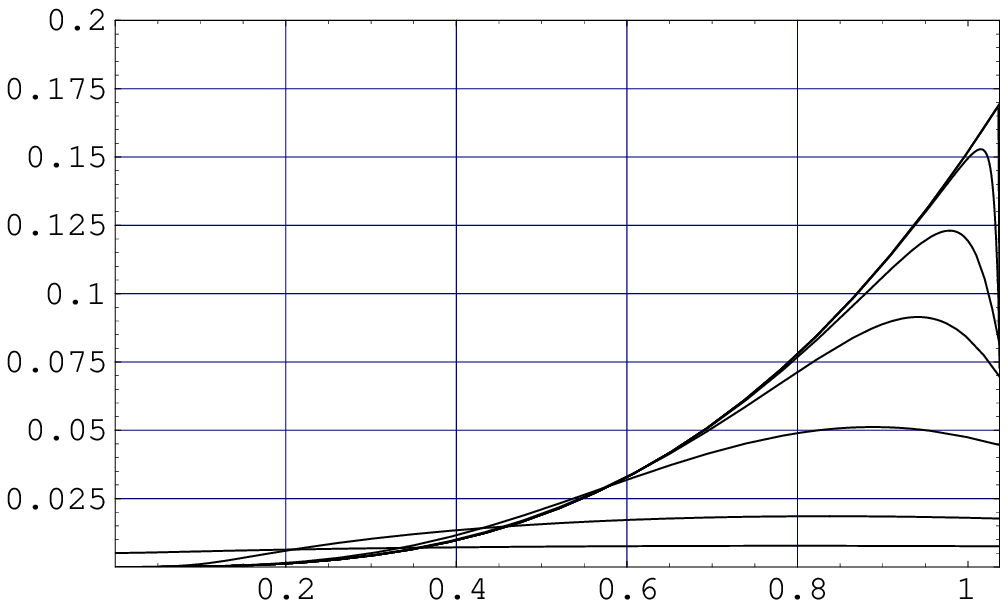}
\hspace{-1cm}
\end{raggedright} 
\centerline{$\lambda \; [\mbox{GeV}]$}
\flushleft \vspace{-4.6cm}\hspace{-0.2cm} {$V_4^{-1}\rho \; [\mbox{GeV}^3]$}
\vspace{+4.0cm} 
\hspace{0.6cm}
\caption{\label{fig:free}The spectrum for different values of
  $m_\Lambda$. As $m_\Lambda$ approaches infinity the curve resembles
  more and more the free spectrum with a sharp cutoff at
  $\Lambda$. The spectra are plotted for
  $m_\Lambda=.228,\;.5,\;1.5,\;4,\;10,\;50,\;2000$ GeV. The free
  spectrum differs from the spectrum with $m_Lambda=2000$ only for
  values of $\lambda$ very close to $\Lambda$.}
\end{center}
\end{figure}

We have presented a method to obtain the spectral density of the Dirac
operator for the linear $\sigma$-model in the large $N_c$
limit. Because of asymptotic freedom in QCD, we expect
$\rho(\lambda)\propto\lambda^3$ for asymptotically large eigenvalues,
since for a free theory the spectral density is only determined by the phase
space. This situation of a non interacting theory can be tested within
our approach in the limit $m_\Lambda\rightarrow\infty$. In this case the NJL
four fermion interaction $G \propto 1/m_\Lambda^2$ \cite{Meyer:2001zp}
vanishes. As shown in fig. (\ref{fig:free}) the resulting spectral
density becomes the density of a free theory. For
$\lambda\geq\Lambda$ the phase space is cut off sharply.
\begin{figure}[h!]
\label{fig:cutoffs}
\begin{center}
\begin{raggedright}
\hspace{1cm}
\epsfig{file=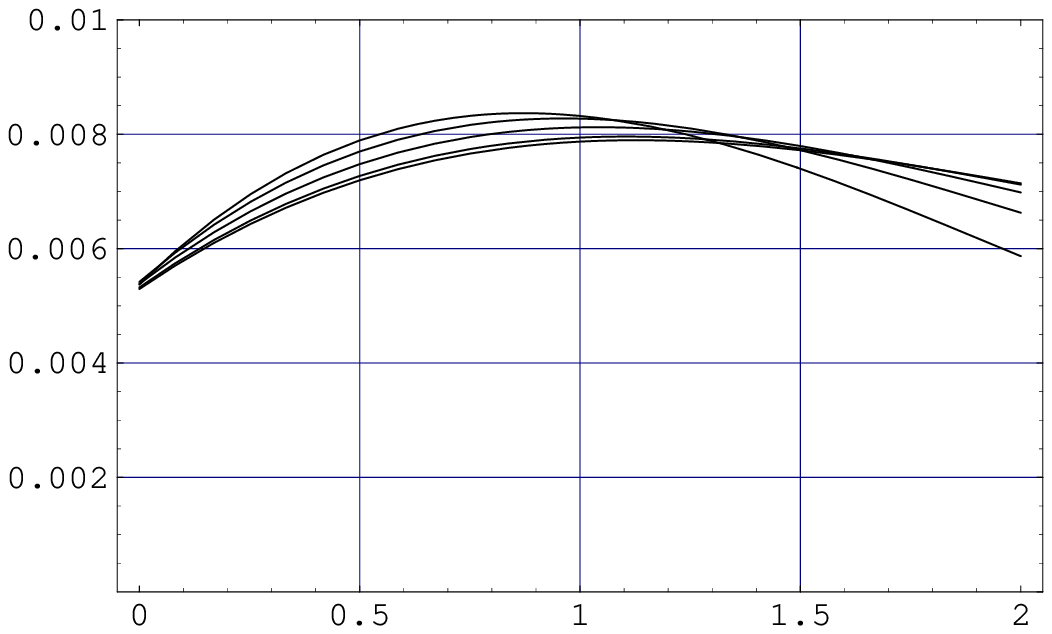}
\hspace{-1cm}
\end{raggedright} 
\centerline{$\lambda \; [\mbox{GeV}]$}
\flushleft \vspace{-4.7cm}\hspace{-0.2cm} {$V_4^{-1}\rho \; [\mbox{GeV}^3]$}
\vspace{+4.1cm} 
\hspace{0.6cm}
\caption{The spectral density for different regulator functions
  $f^{(n)}_k$. The uppermost curve corresponds
  to n=2. The others are respectively $n=3,\;5,\;10,\;15$, where the
  last two are already nearly equal.} 
\end{center}
\end{figure} \\
If the interaction strength $G$ is turned on by a finite $m_\Lambda$
one sees that quark-quark interactions lead to a level repulsion
increasing the density for small $\lambda$ and pushing some strength
above 1 GeV.
Varying $n$ and thereby the shape of the cutoff functions
$f^{(n)}$ in eq. (\ref{eq:cutoffs}) we adjust the cutoff parameters
$m_\Lambda$ and $\Lambda$ in such a way that $f_\pi$ and $M_q$ are
kept constant. This way we can test the sensitivity of the spectral
density to the cutoff function. We see that below 1 GeV the choice of
the cutoff function leads to stable results within $10 \%$. \\
In the realistic case of QCD we expect that gluonic interactions above
$\Lambda$  influence the spectrum nearby and below. 
Therefore, the decrease of the
spectral density $\rho(\lambda)$ above $\lambda \approx 0.8$ GeV is
probably an artifact of the model space, which is resticted by the
soft cutoff. \\
The interactions in the strong coupling regime of QCD induce level
repulsion. Higher modes are strongly suppressed, while the increase of
the average eigenvalue density for smaller eigenvalues drives the
chiral condensate which appears on the low energy
end of the spectrum. It seems that level repulsion and condensation are
the dominating effects for the Dirac operator spectral density in the
strong coupling regime of QCD.

\section{Conclusions}

In this paper we have presented a new method to obtain the Dirac
spectrum with renormalization group flow equations. We have
used proper time flow equations for the linear $\sigma$-model with
quarks of two flavors. In the large $N_c$ approximation we have found a simple
connection between the meson potential $U_k(\Phi)$ and the chiral
condensate which allowed us to calculate the Dirac spectrum directly
from the full effective meson potential $V(\Phi,z)$.
We have recovered the Banks--Casher relation for small
eigenvalues. The next to leading order coefficient is 87 MeV.\\
The calculation of $\rho(\lambda)$ should be valid up to several
hundred MeV. Thus our
approach allows to enter a new regime in the description of the Dirac
spectrum which may complement other results from chiral
perturbation theory and chiral random matrix theory. Compared to a
free Dirac theory we have found the eigenvalue density suppressed for
large eigenvalues and increased for small eigenvalues.
All results are obtained in the infinite volume limit. Since the
Schwinger proper time cutoff functions are simple to handle, the
same calculation can be done for finite Euclidean space time volumes,
where the evolution of the flow equations enters decisively. Such a
calculation could then be compared to QCD lattice
computations \cite{Damgaard:2001xr}. Further we plan to extend the
calculation beyond the large $N_c$ approximation. The renormalization
group flow equations of the linear $\sigma$-model with quarks provide
a framework to include meson loops in the NJL-model without an
additional new cutoff \cite{future}.

\section*{Acknowledgments}
We would like to thank J. Meyer, S. Meyer, H.A. Weidenmueller, M.A. Stephanov, G. Akeman and M.A. Nowak for stimulating discussions.

\end{document}